\documentclass[prb,preprint,amsmath,amssymb,showpacs]{revtex4}
\usepackage[english]{babel}
\usepackage[applemac]{inputenc}
\usepackage{latexsym}

\newcommand{\abs}[1]{\ensuremath{\left| #1 \right|}}    

\newcommand{\cl}{TiOCl}
\newcommand{\bro}{TiOBr}
\newcommand{\wn}{cm$^{-1}$}
\newcommand{\refl}{R($\omega$)}
\newcommand{\sig}{$\sigma_1(\omega)$}

\usepackage{hyperref}
\usepackage{graphicx}

\begin{document}

\title {Analysis of the phonon spectrum in the titanium oxyhalide TiOBr}

\author {G. Caimi }
\author {L. Degiorgi}
\affiliation{Laboratorium f\"ur Festk\"orperphysik, ETH
Z\"urich,
CH-8093 Z\"urich, Switzerland}\

\author {P. Lemmens}
\affiliation{Max Planck Institute for Solid State Research, Heisenbergstr.
1, D-70569 Stuttgart, Germany}\

\author {F.C. Chou}
\affiliation{Center for Materials Science and Engineering, M.I.T.,
Cambridge, MA 02139, U.S.A.}\

\date{\today}

\begin{abstract}
We present the electrodynamic response of TiOBr, which undergoes a
transition to a spin-singlet ground state for temperatures below
T$_{C1}=28$ K. The temperature evolution of the phonon anomalies indicates,
like in \cl, an extended fluctuation regime, extending well above T$_{C1}$. At low frequencies, the spectral weight is progressively suppressed by
decreasing the temperature, suggesting the formation of a spin-gap and a considerable  electron-phonon  coupling. A comparison of the two
oxyhalides shows a weaker interplane coupling and reduced dimensionality
for TiOBr that enhances fluctuations and shifts the ordering temperatures
to lower values. In the fluctuation regime the temperature dependencies of
some phonon parameters track the temperature evolution of the magnetic
susceptibility.
\end{abstract}


\maketitle

\section{Introduction}
The research on two-dimensional (2D) high temperature superconductors raised
considerable interest in related low dimensional transition metal oxides.
The aim is to understand the interplay of topological aspects, strong
electronic correlation and magnetism in low dimension. As quantum magnetism
with spin-$\frac{1}{2}$ is characterized by strong fluctuations and
suppression of long range magnetic order\cite{lem} other ground states with
exceptional properties may be realized. Beside Cu$^{2+}$ in 3$d^9$
configuration with a hole in the $e_{g}$ orbitals, S$=\frac{1}{2}$ quantum
magnets are achieved with Ti$^{3+}$ and V$^{4+}$, namely in the 3$d^1$
configuration and one single $d$ electron occupying a $t_{2g}$ orbital. In
addition, the smaller Jahn-Teller coupling of $t_{2g}$ with respect to
$e_{g}$ orbitals allows in certain cases an additional dynamics of these
orbital states. Initially, the TiOX compounds, with X=Cl and Br, were
considered as 2D antiferromagnets and candidates for resonance valence bond
ground states\cite{bey}. Recently, however, new experimental findings in
TiOCl jeopardize this picture and point toward a one dimensional character
of the electronic system. LDA+U calculations predict an ordering of the
$t_{2g}$ orbitals leading to a quasi 1D antiferromagnetic
spin-$\frac{1}{2}$ chain at low temperature\cite{seidel}. Nonetheless,
pronounced phonon anomalies observed in Raman and infrared (IR)
spectroscopy suggest a gradual cross-over of the magnetic correlations from
1D at low temperature to 2D at higher temperature, due to a change of the
$t_{2g}$ admixture\cite{lmio,mio}. These experimental findings are
supported by a recent band structure calculation using frozen phonons that
shows a hybridization of oxygen and chlorine states and an admixture of the
$d_{xz}$ and $d_{yz}$ orbitals induced by
the local distortions.  These latter effects can contribute to a 2D character of the
ground state \cite{valenti03}.

The thorough analysis of the phonon anomalies in \cl~(Refs.
\onlinecite{lmio} and \onlinecite{mio}) evidenced a narrowing of the phonon
modes and depression of their spectral weight ($SW$) for temperatures below
T$^* = 135$ K, which was associated with a pseudo spin-gap phase, first
identified by NMR experiments\cite{imai}. Using this depletion of $SW$
observed in both Raman and IR spectroscopy a spin-gap of $2\Delta\approx
430$~K has been inferred  (Refs. \onlinecite{lmio} and
\onlinecite{mio}). The most interesting aspect of \cl~is the pseudo
spin-gap phase at intermediate temperature and the interplay of phonon with
orbital degree of freedoms. At low temperatures, T$<$T$_{C1}=67$ K, TiOCl
shows a more conventional spin-Peierls-like (SP) behavior.

In order to broaden our knowledge on titanium oxyhalides and trying to
generalize the picture already drawn\cite{lmio,mio} for TiOCl, we have
investigated TiOBr by infrared spectroscopy. In TiOBr, due to the larger
ionic radius of Br$^{-}$, a weaker interplane coupling is expected that
should lead to even stronger fluctuations. In this paper we will provide a
complete set of optical data, followed by a detailed analysis of the phonon
spectrum of TiOBr. A comparison with the parent compound TiOCl is performed
as well. The manuscript is organized as follow: first we briefly describe
the experimental results. We then present the analysis of the data and the
related fitting procedure, which allows us to extract the relevant
informations about the ground state, the spin-gap opening and the fluctuation
effects in the TiOX systems.

\section{Experiment and Results}

\begin{figure} [!h]
\begin{center}
  \resizebox*{9.0 cm}{!}{\includegraphics{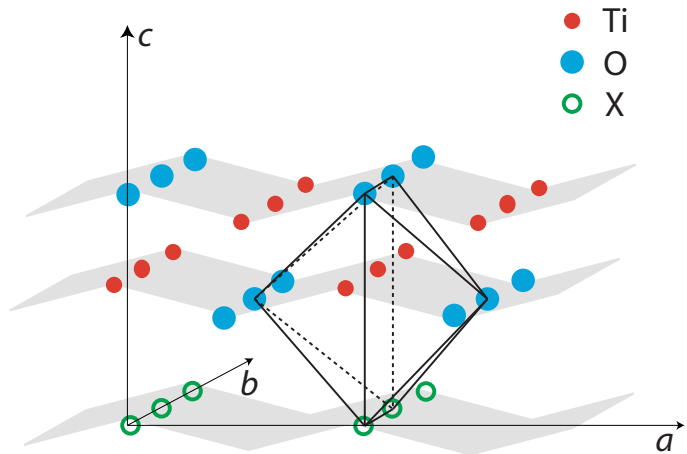}}
\caption{(Color online) Schematic representation of the crystal structure
of the TiOX compounds. The octahedra surrounding each Ti ions is traced
out.} \label{structure}
\end{center}
 \end{figure}

The \bro~single crystals were synthesized by vapor-transport techniques
from TiO$_2$ and TiBr$_3$ as reported in Ref. \onlinecite{schaefer}. The
structure of the oxyhalogenide TiOBr (Fig. \ref{structure}) has FeOCl as
archetype and is thus formed by a double layer of Ti$^{3+}$O$^{2-}$
intercalated by a Br$^{-}$ bilayer.

The Br$^{-}$ bilayers mediate only weak van der Waals interactions
between successive Ti$^{3+}$O$^{2-}$ bilayers, inside which Ti and O form a
buckled double plane\cite{bey}. Each Ti ion is surrounded by a distorted
octahedron of O and Br ions. The TiO$_4$Br$_2$ octahedra have the apexes
along the $a$ axis occupied by two O ions. The sides parallel to the  $b$
axis are formed either by two O or by two Br (Fig. \ref{structure}). The
important exchange path in the electronic structure is  the direct $t_{2g}$
orbital overlap, rather than the super-exchange via oxygen\cite{seidel}. The
$t_{2g}$ $d_{xy}$ orbitals form a linear chain running along the $b$ axis,
linking Ti ions in the same plane\cite{seidel}. The $d_{xz}$ orbitals are
rotated by 45$^\circ$ so that two of the lobes point toward the Ti atoms of
the neighbor layer, forming a zig-zag chain along the $a$ axis.

The magnetic susceptibility $\chi$(T) (Ref. \onlinecite{choup}) displays a broad maximum
at high temperatures for both TiOX compounds. At T$_{C1}\sim67$ K and
T$_{C2}\sim92$ K a sharp drop and a kink is observed in TiOCl. In TiOBr
these anomalies are shifted to lower temperatures, T$_{C1}\sim28$ K and
T$_{C2}\sim47$ K, respectively. For T$<$T$_{C1}$, the TiOX compounds have a
non magnetic ground state, which is associated to the opening of a spin-gap
and a related dimerization of the chain of $d_{xy}$ orbitals. This scenario
has some similarities to a spin-Peierls transition\cite{seidel}. The reduced T$_{C}$ in \bro~with respect to \cl~may be phenomenologically explained as a consequence of  the increased distance between the Br and TiO bilayers, which reduces the interplane coupling and enhances the 2D character of \bro. At lower dimensions, the more pronounced  fluctuations hamper the formation of long-range order.

We have measured the optical reflectivity R($\omega$) in a broad spectral
range (30-10$^5$ \wn) as a function of temperature ranging from 10 to 300 K
and at selected magnetic fields  0-7 T. As in the Cl compound, no magnetic
field dependence is observed in \bro~at any temperature. We will focus our
attention on the temperature dependence only. In the far- (FIR) and
mid-infrared (MIR) spectral range (i.e., 30-5000 \wn), the \refl~spectrum
was measured with a Fourier interferometer with the sample placed inside a
magneto-optical cryostat equipped with appropriate optical windows. The
visible and UV spectral ranges were measured with a home made spectrometer
based on a Zeiss monochromator and a commercial McPherson  spectrometer,
respectively. Light was linearly polarized along the chain $b$ axis and the
transverse $a$ axis.  The same care, as  reported in Ref. \onlinecite{mio},
was applied in order to avoid leakage effects of the polarizer and  to
assure that no undesired projections of the light polarization along any
crystallographic direction occur in the present experiment.

\begin{figure} [!h]
\begin{center}
    \resizebox*{9.0 cm}{!}{\includegraphics{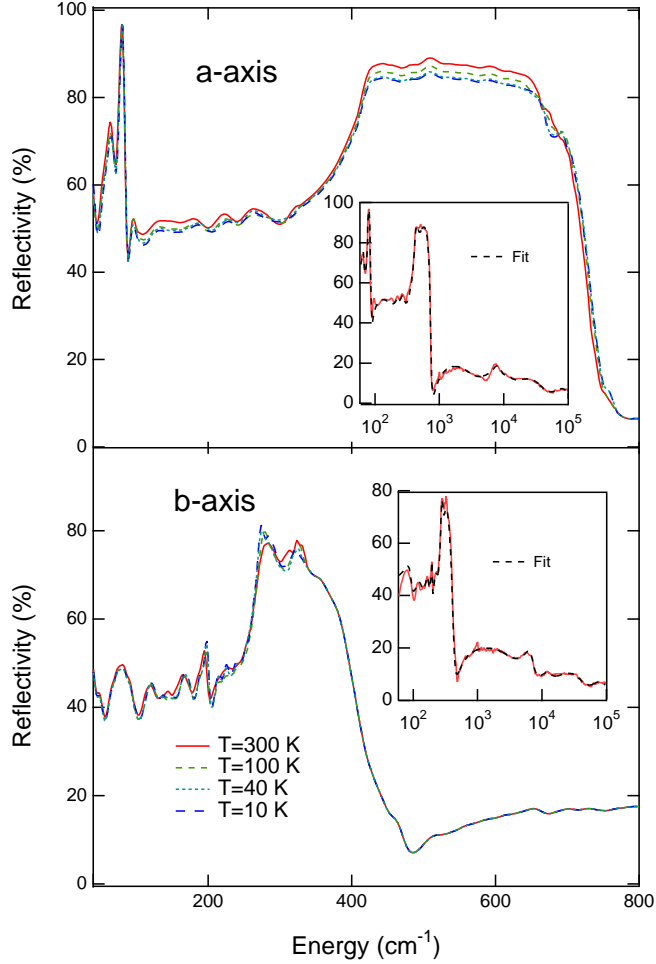}}
\caption{(Color online) \refl~of \bro~at selected temperatures for light
polarized along the $a$ axis (upper panel) and the $b$ axis (lower panel).
The insets show the spectra in the whole measured energy interval and their
Lorentz-Drude fit (see text). } \label{ref}
\end{center}
\end{figure}

\begin{figure} [!h]
\begin{center}
    \resizebox*{9.0 cm}{!}{\includegraphics{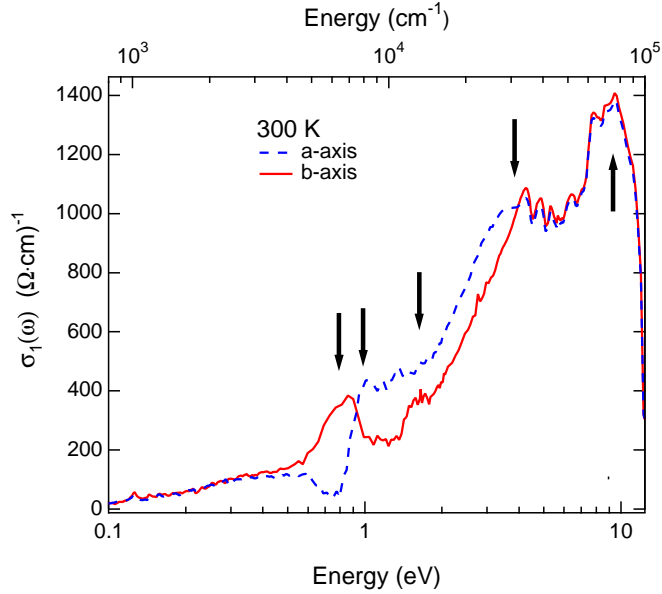}}
\caption{(Color online) High frequency part of  \sig~at 300 K for light
polarized along the $a$  and $b$ axis. The arrows indicate the
characteristic absorption features (see text).} \label{sig}
\end{center}
\end{figure}

Figure \ref{ref} displays the temperature dependence of \refl~in the
infrared spectral range. At high frequencies the spectra are temperature
independent. The upper panel shows the \refl~spectra measured with light
polarized along the transverse $a$ axis, while the lower one shows the
spectra taken with light polarized along the chain $b$ axis. The samples
were too thin to allow an investigation of  the electrodynamic response
along the c-axis. Comparing both polarizations, one can appreciate the
anisotropy of the reflectivity spectrum. The insets of Fig.
\ref{ref} present the whole \refl~spectra up to the UV frequencies. The
\refl~spectra, suitably extrapolated outside the measured frequency range
(for  details see Refs. \onlinecite{wooten} and \onlinecite{Dressel}), can
be used  to calculate the  optical functions, such as the real part
$\sigma_1(\omega) = Re (\tilde{\sigma}(\omega))$ of the optical
conductivity through Kramers-Kronig (KK) transformation. In Fig. \ref{sig}
we plot the high frequency part  of \sig~for both polarizations, which
highlights the absorption spectrum associated with the electronic interband
transitions. The FIR part of \sig~is shown in Figs. \ref{compA} and
\ref{compC}.  The \sig~spectra turn out to be  polarization independent
above $\sim 4\cdot10^4$ \wn.

\section{Discussion}

\begin{figure} [!h]
\begin{center}
    \resizebox*{9.0 cm}{!}{\includegraphics{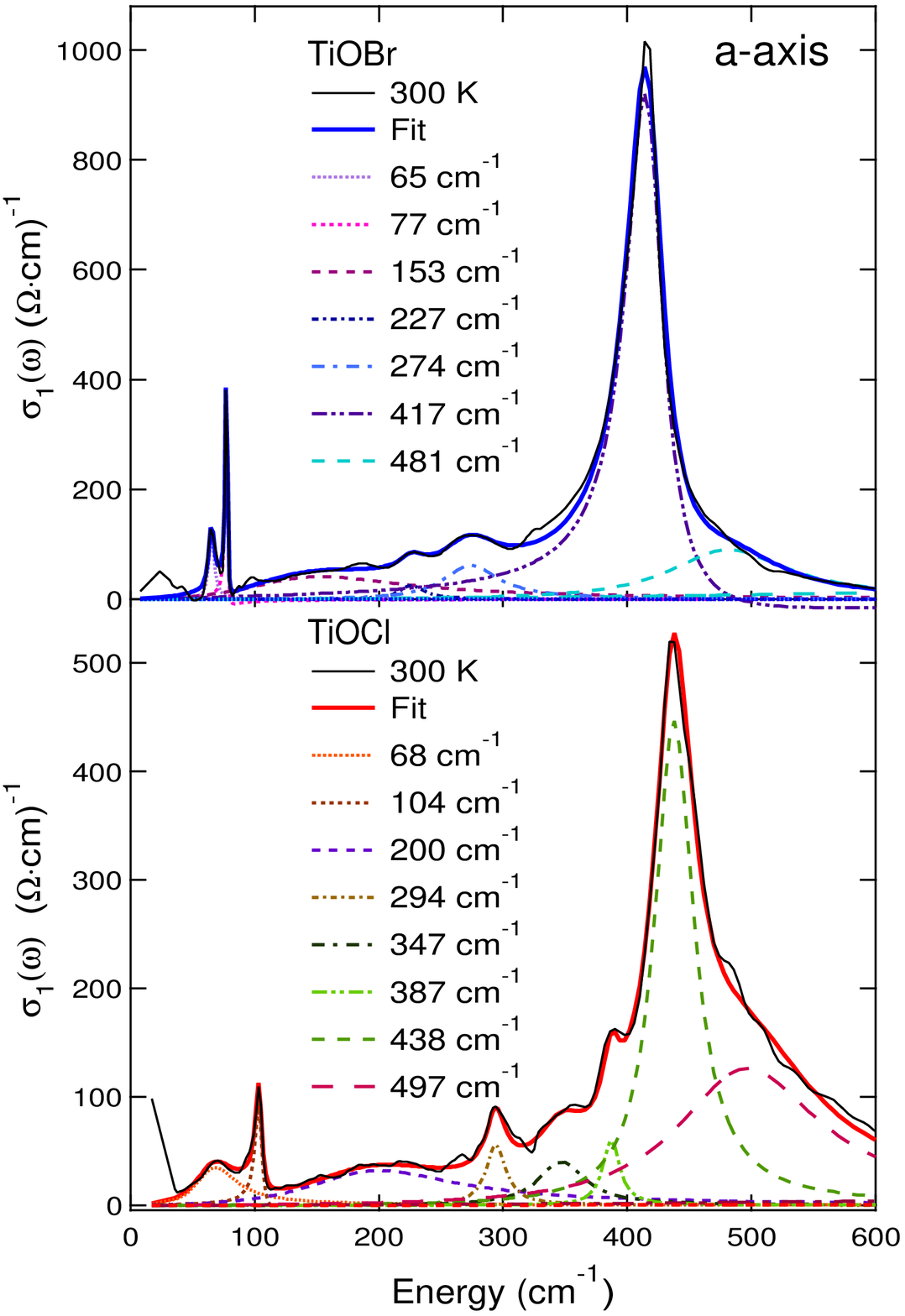}}
\caption{(Color online) The optical conductivity of \bro~in FIR along the
$a$ axis is displayed in the upper panel. The total fit and its components,
identified in the legend by their respective resonance frequency, are also shown \cite{tab}. \sig, the total fit and its
components for \cl~(Ref. \onlinecite{mio}) are shown in the lower panel
for the purpose of comparison.} \label{compA}
\end{center}
\end{figure}

\begin{figure} [!h]
\begin{center}
    \resizebox*{9.0 cm}{!}{\includegraphics{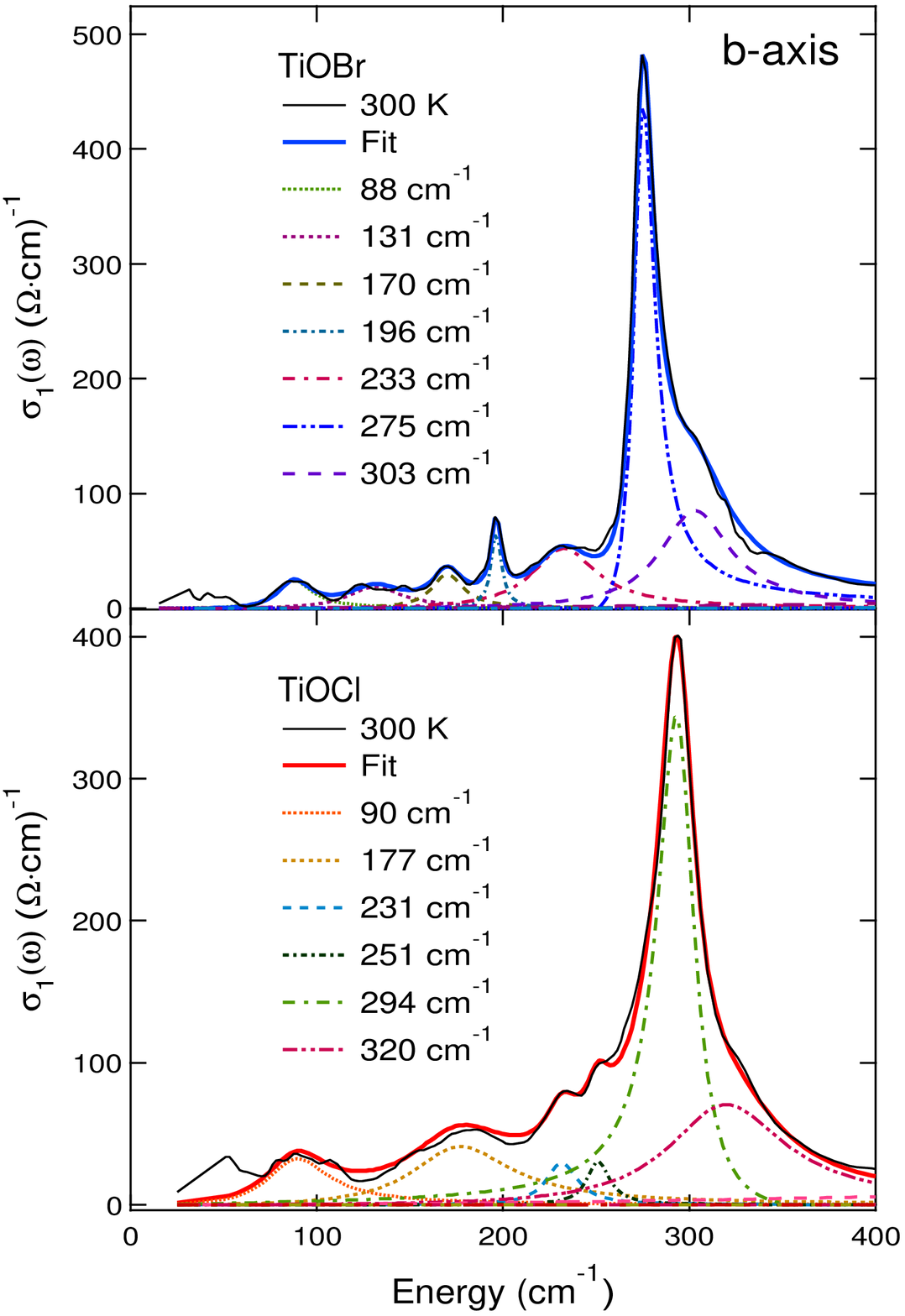}}
\caption{(Color online) The optical conductivity of \bro~in FIR along the
$b$ axis is displayed in the upper panel. The total fit and its components,
identified in the legend by their respective resonance frequency, are also shown \cite{tab}. \sig, the total fit and its
components for \cl~(Ref. \onlinecite{mio}) are shown in the lower panel
for the purpose of comparison.} \label{compC}
\end{center}
\end{figure}

The high frequency part of  \sig, reflecting the electronic interband
transitions, is characterized by  four main absorptions (evidenced by the
arrows in Fig. \ref{sig}). The first two are at $\sim 0.8-1$ eV and
$\sim1.5$ eV, the third  at $\sim3.8$ eV and the fourth at $\sim9.3$ eV.
These energies are comparable with those observed in \cl~(Refs.
\onlinecite{lmio} and \onlinecite{mio}). Recent LDA+U calculations on \cl
\cite{seidel}, using a full-potential LMTO method predict a split-off of
the one dimensional $t_{2g}$ bands creating an insulating charge gap of
about $\sim1$ eV. This seems to occur in \bro, as well. The same band
structure calculation indicates interband transitions between the O and  Cl
p-level and the Ti d-levels at energies between 4 and 7 eV. Since the
atomic distances within the $ab$ plane vary only slightly between the two
compounds \cite{bey} and neglecting the interplane coupling, we can assume
that the TiOX compounds have a similar electronic band structure.
Therefore, it is of no surprise that both titanium oxyhalides display similar absorption features\cite{mio}.

Several absorptions characterize \sig~in the FIR spectral region and they
are listed in Tables \ref{tabphA} and \ref{tabphC} for the $a$  and $b$
axis, respectively. Along the $a$ axis (Fig. \ref{compA}), a strong
absorption is seen at 417 \wn, with a broad high frequency tail defining a
shoulder at about 481 \wn. At low frequencies, we detect a sharp absorption
at 77  \wn~and a small one at 65 \wn. Along the $b$ axis (Fig. \ref{compC})
a strong absorption is observed at 275 \wn, with a shoulder at 303
\wn~defining its high frequency tail.

In the lower panels of  Figs. \ref{compA} and \ref{compC},  \sig~of TiOCl
has been added for the purpose of comparison \cite{mio}. One notes the overall 
similarity between the spectra of both compounds. Nonetheless, in \bro~a
generalized red-shift of the phonon spectrum with respect to \cl~is observed as may be expected
when replacing Cl by a heavier element such as Br. There are, however, some
differences, like the number and shape of the absorption modes between the
two TiOX compounds. For example,  the strong absorption peak at 417 \wn~in
\bro~along the $a$ axis shows a quite pronounced asymmetry,  not observed
at the 438 \wn~absorption of \cl.  Along the $b$ axis, the strong
absorption  at 275 \wn and the weaker modes at 233 \wn~are well
distinguished in \bro, while in  \cl~one finds  a unique asymmetric mode at
294 \wn. Similarly, the mode at 387 \wn~is well resolved in \cl~but not in \bro.

\begin{table}[!h]
\caption{Resonance frequencies of the FIR absorptions in \bro~(i.e., peaks
in \sig) along the $a$ axis, determined by the  Fano approach (eq. (\ref{fan})). The corresponding frequencies for TiOCl (Ref.
\onlinecite{mio}) are listed for comparison. The
table also reports the ratio of the resonance frequencies between the Cl
and Br compound, to be compared with eq. (\ref{far}) and (\ref{mar}). The
bold frequencies refer to the $B_{3u}$ modes. }

\label{tabphA} \vspace{0.5cm} \centering\begin{tabular}{ccccccccc}
\hline\hline
\multicolumn{9}{c}{{\bf $a$ axis}}\\
\hline\hline
\bro&65&\bf{77}&153&227&274&&\bf{417}&481\\
\hline
\cl&68&\bf{104}&200&294&347&387&\bf{438}&497\\
\hline
$\omega_{0_{Br}}/\omega_{0_{Cl}}$&0.96&\bf{0.74}&0.77&0.77&0.79&&\bf{0.95}&0.97\\
\hline\hline
\end{tabular}
\end{table}

\begin{table}[!h]
\caption{Resonance frequencies of the FIR absorptions in \bro~(i.e., peaks
in \sig) along the $b$ axis, determined by the  Fano approach (eq. (\ref{fan})). The corresponding frequencies for TiOCl (Ref.
\onlinecite{mio}) are listed for comparison. The
table also reports the ratio of the resonance frequencies between the Cl
and Br compound, to be compared with eq. (\ref{far}) and (\ref{mar}). The
bold frequencies refer to the $B_{2u}$ modes. } \label{tabphC} \vspace{0.5cm}
\centering\begin{tabular}{cccccccc} \hline\hline
\multicolumn{8}{c}{{\bf $b$ axis}}\\
\hline\hline
\bro&88&\bf{131}&170&196&233&\bf{275}&303\\
\hline
\cl&90&\bf{177}&231&251&&\bf{294}&320\\
\hline
$\omega_{0_{Br}}/\omega_{0_{Cl}}$&0.97&\bf{0.74}&0.74&0.78&&\bf{0.94}&0.95\\
\hline\hline
\end{tabular}
\end{table}

The TiOX compounds are characterized by  the space group $P_{mmn} (59,
D_{2h})$  at room temperature. The infrared active phonons as well as the
eigenvectors for the atomic displacements have been established with shell
model calculations. The results of this simple phonon modes calculations are reported in
Ref. \onlinecite{mio}. Two $B_{3u}$ modes polarized along the $a$ axis, two
$B_{2u}$ along the $b$ axis and two $B_{1u}$ along the c-axis were
predicted.   We can  describe the phonon modes of the Br compound using our
knowledge on the Cl compound\cite{mio} and taking into account the
renormalization of the phonon frequencies due to the corresponding ion
reduced mass and as consequence of the change in the lattice coupling,
which follows from the relative volume variation of the unit cell when
substituting Br with Cl.

Within the linear harmonic approximation, the shift of the eigenfrequencies
of the phonons is obtained by a  renormalization of the oscillator strength
constant $f$, when Br and Cl ions are not involved in the oscillatory
displacements:
\begin{equation}
\label{fr}
\omega_{0_{Br}}=\sqrt{\frac{f_{Br}}{m}}=\sqrt{\frac{f_{Cl}\cdot\frac{f_{Br}}{f_{Cl}}}{m}}=\omega_{0_{Cl}}\cdot\sqrt{\frac{f_{Br}}{f_{Cl}}}
\end{equation}

We may estimate $\sqrt{\frac{f_{Br}}{f_{Cl}}}$ for both directions ($a$ and
$b$ axis) from the IR optical active phonons, whose frequency is predicted
to be  independent from  $m_{Br/Cl}$ (Ref. \onlinecite{mio}):
\begin{equation}
\label{far}
\sqrt{\frac{f_{Br}}{f_{Cl}}}=\left \{ \begin{array}{ll}
\frac{\omega_{0_{Br}}}{\omega_{0_{Cl}}}=\frac{417}{438}=0.9521& \textrm{for {\bf \textit{a}-axis}}\\
\frac{\omega_{0_{Br}}}{\omega_{0_{Cl}}}=\frac{275}{294}=0.9354& \textrm{for {\bf \textit{b}-axis}}.\\
\end{array}
\right.
\end{equation}
On the other hand, for the IR phonons, where  the mass of the  Cl/Br ions is involved, one should consider  also the renormalization due to the reduced mass.
\begin{equation}
\label{fmar}
\omega_{0_{Br}}=\sqrt{\frac{f}{\mu_{Br}}}=\sqrt{\frac{f_{Cl}\cdot\frac{f_{Br}}{f_{Cl}}}{\mu_{Cl}\cdot\frac{\mu_{Br}}{\mu_{Cl}}}}=\omega_{0_{Cl}}\cdot\sqrt{\frac{\mu_{Cl}}{\mu_{Br}}}\cdot\sqrt{\frac{f_{Br}}{f_{Cl}}}
\end{equation}

According to the eigenvector illustrated in Fig. 3 of Ref.
\onlinecite{mio}, one  has one atom of Ti and O for each unit cell, moving
together against the Br atom. Therefore, the $\rm{Ti} + \rm{O}$ ensemble
has a resulting atomic mass $m=m_{Ti}+m_{O}= 63.87~ \rm{au}$. The reduced
mass $\mu$ of the eigenmode is then:
\begin{eqnarray}
\label{redm}
\mu_{Cl}&=&\frac{m}{1+\frac{m}{m_{Cl}}}=22.80~ \rm{au}\nonumber\\
\mu_{Br}&=&\frac{m}{1+\frac{m}{m_{Br}}}=35.50 ~\rm{au}.
\end{eqnarray}
for the Cl and  Br compound, respectively. Inserting these reduced masses
in  eq. (\ref{fmar}), the expected red-shift of  the IR active phonons in
\bro~can be estimated as:
\begin{equation}
\label{mar}
\omega_{0_{Br}}=\omega_{0_{Cl}}\cdot\sqrt{\frac{\mu_{Cl}}{\mu_{Br}}}\cdot\sqrt{\frac{f_{Br}}{f_{Cl}}}=\omega_{0_{Cl}}\cdot \left \{ \begin{array}{ll}
0.7630&  \textrm{for {\bf \textit{a}-axis}}\\
0.7496&  \textrm{for {\bf \textit{b}-axis}}.\\
\end{array}
\right.
\end{equation}

This  simple approach accounts very well for the generalized red-shift of
the phonon spectrum in \bro~with respect to \cl. There is an excellent
agreement between the scaling following eq. (\ref{far}) and (\ref{mar}) and the
measured red-shift of the $B_{2u}$ and $B_{3u}$ modes, as may be seen by
the ratios $\omega_{0_{Br}}/\omega_{0_{Cl}}$ in Tables \ref{tabphA} and
\ref{tabphC}. Applying this argument to the eigenfrequencies, calculated
for the $B_{2u}$ and $B_{3u}$ modes of \cl~(Ref. \onlinecite{mio}), we can
anticipate the expected (theoretical) eigenfrequencies for \bro. These
results are listed in Table \ref{tabphonon}.  The agreement with the
experimentally determined values (Tables \ref{tabphA} and \ref{tabphC}) is
rather compelling.

\begin{table}[!h]
\caption{ IR phonon mode frequencies for \cl~after the shell model
calculation reported in Ref. \onlinecite{mio} and the  estimated ones for
\bro~using the renormalization factors calculated by eq. (\ref{far}) and
(\ref{mar}). }\label{tabphonon} \vspace{0.5cm}
\centering\begin{tabular}{ccccccc} \hline\hline
&~~~&\multicolumn{2}{c}{$B_{2u}$~({\bf $b$ axis})}&~~~&\multicolumn{2}{c}{$B_{3u}$~({\bf $a$ axis})}\\
\hline
\bro~Theory&&148&311&&69&410\\
\hline
\cl~Theory&&198&333&&91&431\\
\hline
\hline
\end{tabular}
\end{table}

Also in \bro~more modes are detected than expected by symmetry\cite{mio}.
Nonetheless, we can extend the above analysis to these additional modes, as
well. In Tables \ref{tabphA} and  \ref{tabphC}, we have reported all mode
frequencies measured along the $a$  and $b$ axis, which  are compared with
those of \cl~(Ref. \onlinecite{mio}). It is worth noting that the modes
ranging from 77 up to 347 \wn~along the $a$ axis shift upon Br substitution
of $\sim 77-79\%$, indicating that for these modes the Br mass plays an
active role. The high frequency tail of the strong phonon as well as  the lowest
phonon at 65 \wn, scale by $\sim 96-97\%$, suggesting that  only the
renormalization of the lattice strength plays here a relevant role. The mode
scaling  along the $b$ axis is similar to that of the $a$ axis. Those modes
enclosed between the two predicted IR  phonons  (131-275 \wn) show a
softening of $\sim 74-78\%$, face a prediction of $\sim 75\%$. The shoulder at high frequency of the 
 $B_{2u}$ mode, generated by the displacement of the  Ti and
O  ions, and the  lowest phonon at 88 \wn, display a shift of $\sim
95-97\%$, confirming that in these two modes the Br displacement is generally
not relevant. Even though interference effects cannot be excluded a priori, the fact that all FIR absorptions for both polarization
directions scale following eq. (\ref{far}) or eq. (\ref{mar}) supports the
lattice dynamics origin for these excitations in both compounds. Similarly for TiOCl, we believe that in TiOBr as well the effective space group can be used to identify a lower crystallographic symmetry.
Such a lowering in symmetry enlarges the number of infrared-activated
modes (like purely Raman ones), which otherwise would be silent or
even IR-forbidden.

In order  to highlight the temperature dependence of the phonon spectrum,
we apply the phenomenological Fano approach\cite{fano} to the optical
conductivity $\tilde{\sigma}(\omega)$, already introduced in Ref.
\onlinecite{mio} and successfully employed for \cl:
\begin{equation}
\label{fan}
\tilde{\sigma}(\omega)=\sigma_1(\omega)+i\sigma_2(\omega)=\sum_{j}i\sigma_{0j}\frac{(q_{j}+i)^{2}}{i+x(\omega)},
\end{equation}
with $x(\omega)=(\omega_{0j}^{2}-\omega^{2})/\Gamma_{j}\omega$, where
$\omega_{0j}$ is the resonance frequency, $\Gamma_{j}$ is the width (i.e.,
damping) and $\sigma_{0j}=\omega_{pj}^{2}/\Gamma_{j} q_{j}^{2}$ with
$\omega_{pj}$ as the oscillator strength and $q_{j}$ as the so-called
asymmetry factor of the $j$-absorption. Eq. (\ref{fan}) allows us to
describe the asymmetry in the line shape of the absorption features in
\sig, which derives from the interaction of  localized states (i.e., phonon
modes) with a continuum (i.e., the continuum extends over a spectral range at energies below the localized state if $q_j<0$; at energies above the localized state if $q_j>0$). The  fit of the complete \sig~is obtained by
summing over twelve contributions (the seven shown in Figs. \ref{compA} and \ref{compC} and five more for the high frequency spectral range)  for both the $a$  and $b$ axis (the
complete set of fitting parameters is given in  Ref. \onlinecite{tab}). The
fit for the low frequency part of \sig~at 300 K as well as its single
components are shown in Figs. \ref{compA} and \ref{compC}. The reproduction
of the experimental data is astonishingly good and the same fit quality is
obtained at all temperatures. The same set of fit parameters\cite{tab}
allows us to reproduce the measured \refl~spectra with great fit quality as
demonstrated by the insets in Fig. \ref{ref}. Only the seven oscillators with
the lowest energy (Figs. \ref{compA} and \ref{compC}) display  a
temperature dependence and will be here discussed further.

\begin{figure} [!h]
\begin{center}
    \resizebox*{9.0 cm}{!}{\includegraphics{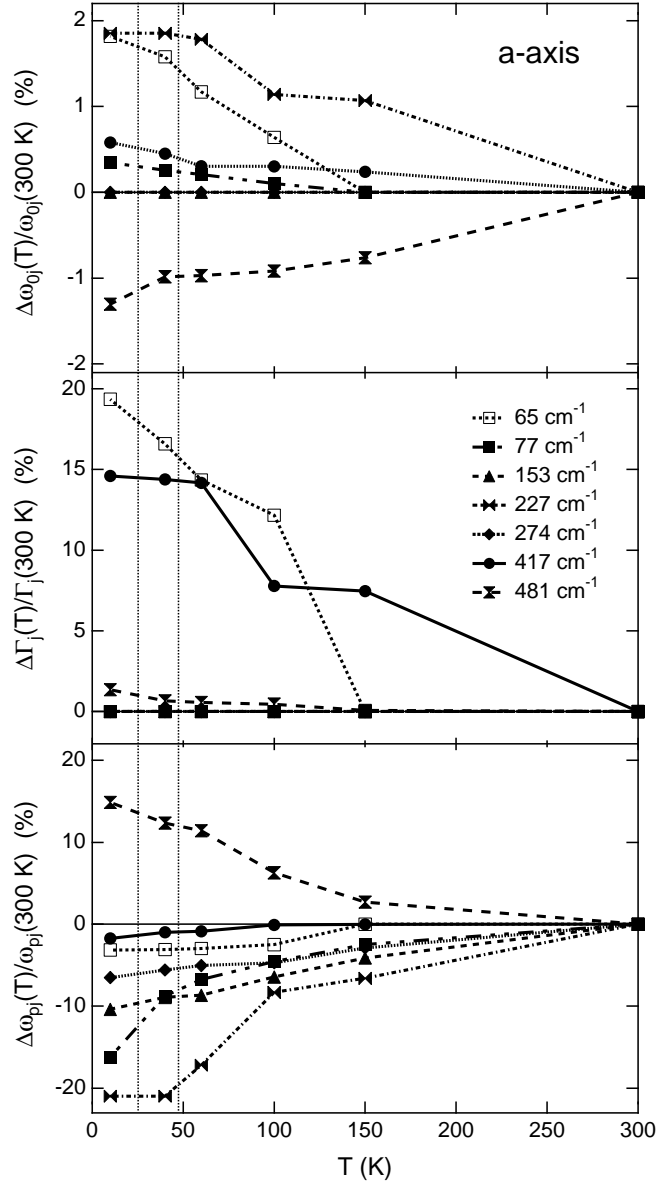}}
\caption{Temperature dependence along the $a$ axis of the
percentage variation with
respect to 300 K (see text) for the resonance frequencies
($\omega_{0j}$), the dampings
($\Gamma_{j}$) and the oscillator strengths ($\omega_{pj}$)
of the phonon modes (identified in the legend by their respective
resonance frequency in \wn). The dotted vertical lines mark the transition temperatures T$_{C1}$ and T$_{C2}$.}
\label{pa}
\end{center}
\end{figure}

\begin{figure} [!h]
\begin{center}
    \resizebox*{9.0 cm}{!}{\includegraphics{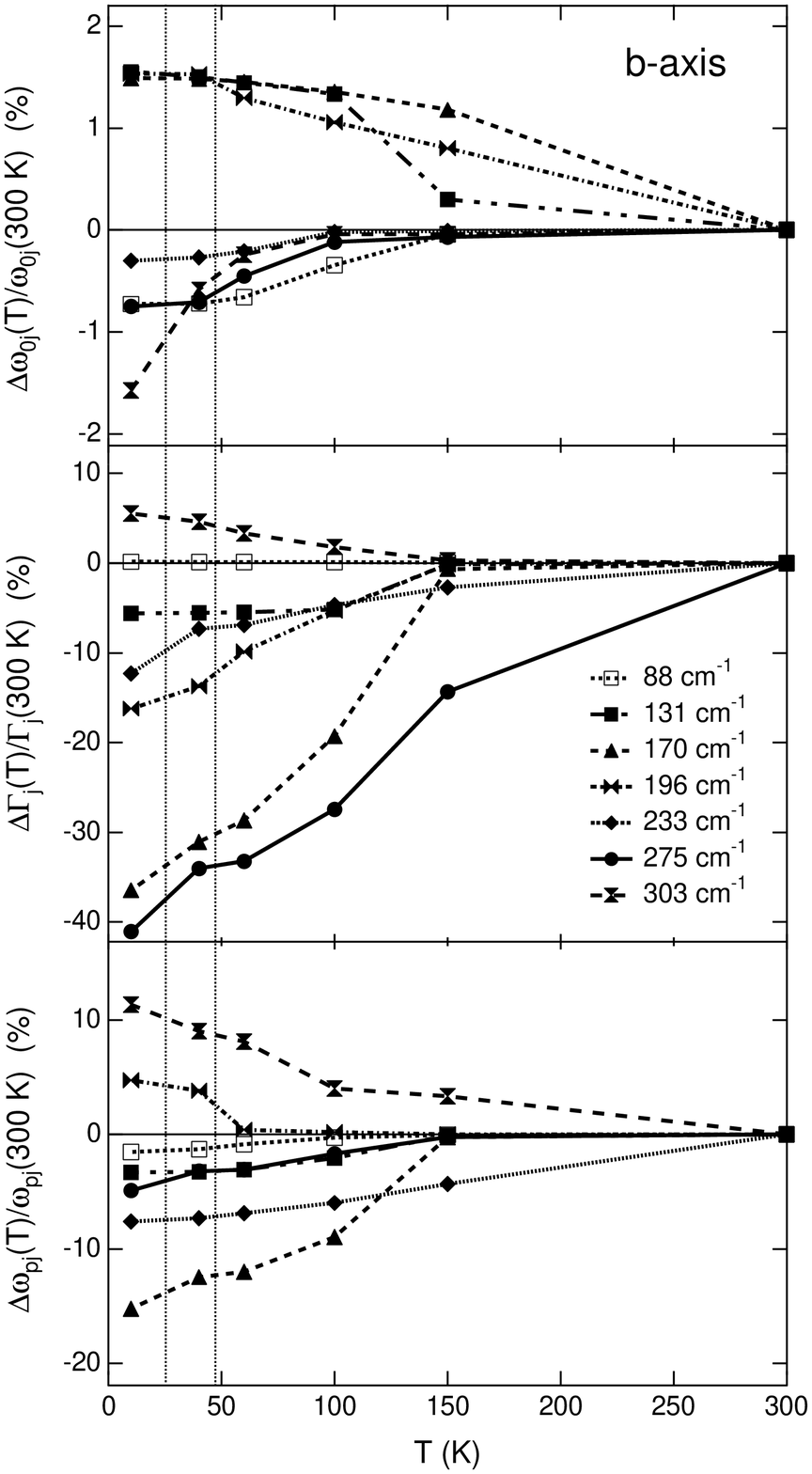}}
\caption{Temperature dependence along the $b$ axis of the
percentage variation with
respect to 300 K (see text) for the resonance frequencies
($\omega_{0j}$), the dampings
($\Gamma_{j}$) and the oscillator strengths ($\omega_{pj}$)
of the phonon modes (identified in the legend by their respective
resonance frequency in \wn). The dotted vertical lines mark the transition temperatures T$_{C1}$ and T$_{C2}$.}
\label{pc}
\end{center}
\end{figure}

The $q$ factor, describing the asymmetry of the phonon at 77 \wn~along the
$a$ axis, increases from   $\sim -8$ at  300 and 150 K to values of about $
-30$ at low temperatures\cite{tab}. This indicates a transition from an
asymmetric mode to a symmetric one. Indeed, the Fano formalism for
$\abs{q}$ factors  larger than ${30}$ is almost equivalent to the (symmetric)
Lorentz model.  The corresponding resonance at 104 \wn~in \cl~is also
asymmetric even though the temperature dependence of its $q$ factor is more
moderate than in \bro. The strong phonon at 417 \wn~along $a$ axis, shows
an asymmetric line shape, which however  does not change in temperature.
Along the $b$ axis, the  peak at 275 \wn~is  asymmetric  with  constant $q$
at all temperatures. Its counterpart in \cl~(i.e., the mode at 294 \wn) displays a
clear crossover from an asymmetric Fano-like  to a Lorentzian line shape
between 200 K and T$_{C2}$. That suggests the suppression of the continuum
interacting with the phonon  at 294 \wn. Since $q<0$ the continuum extends at energies
below the resonance frequency of the phonon. Therefore, its suppression
identifies a characteristic energy scale of $\sim 430$ K. Combining this
finding with  other experimental evidences, like the depletion of spectral
weight at low energies\cite{lmio,mio} and the suppression of the Raman continuum\cite{lmio}, the characteristic energy of about
430 K can be associated to the opening of a spin-gap.  The temperature
independence of the  $q$ factor for the mode at 275 \wn~along the $b$ axis
in \bro~indicates, that the characteristic energy scale for the spin-gap
lies at lower energies than 275 \wn$\approx400$ K.

The temperature dependence of the fit parameters\cite{tab} ($\omega_{0j},
\Gamma_{j}, \omega_{pj}$) is shown in Figs. \ref{pa} and \ref{pc} for both
polarizations as percentage variation with respect to the 300 K data (e.g.,
$\Delta\omega_{0j}$(T)$/\omega_{0j}(300$ K), with
$\Delta\omega_{0j}$(T)$=\omega_{0j}$(T)$-\omega_{0j}(300$~K)). The two critical temperature of the spin system, measured by $\chi$(T) (Ref. \onlinecite{choup} and  \onlinecite{short}), have been traced out with a dotted  vertical line. The overall
temperature dependence of the fit parameters develops in a broad
temperature interval below 150 K, and tends to saturate below 30 K. The
fact, that this temperature interval extends well above T$_{C1}$,
underlines the presence of an extended fluctuation regime. Raman data
confirm such a fluctuation regime in the same temperature
interval\cite{short}. Fluctuation effects  in \mbox{\cl}~have been first
pointed out by NMR measurements\cite{imai}, where the relaxation rate of
$^{47,49}$Ti sites (1/T$_1$T) as a function of temperature, which probes
the spin degree of freedom, shows a maximum at T$^*$. The 1/T$_1$T decrease
below T$^*$ indicates the suppression of low frequency spin fluctuation,
and  below T$_{C1}$, the relaxation rate displays an exponential decay,
signaling thus the opening of a spin gap\cite{imai}. The temperature
evolution of the relaxation rate of $^{35}$Cl develops below $\sim200$ K,
suggesting that the lattice dynamic and the spin degree of freedom
interplay already at high temperatures. Also electron-spin resonance (ESR)
parameters highlight the strong coupling between spin and lattice degree of
freedom, and they shows a progressive evolution in a temperature interval
ranging from 200 K to T$_{C1}$ (Ref. \onlinecite{Kataev}).

The resonance frequency ($\omega _{0j}$) of almost all phonons along the
$a$ axis (upper panel in Fig. \ref{pa})  tends to increase with decreasing
temperature, though moderately, with changes not exceeding  2 \%. Only the
highest mode at 481 \wn~shows a weak softening. Along the $b$ axis, the low
energy phonons display a weak hardening, while the three modes around the
strong absorption  feature peaked at 275 \wn~show a weak softening (upper
panel in Fig. \ref{pc}). A softening of the phonon modes is in principal
expected in models for a conventional spin-Peierls transition \cite{cross},
where the structural deformation is driven by a linear coupling between the
lattice and the magnetic degrees of freedom. However,  since the
dimerization must be related to normal modes away from the zone center, the
softening of one or more modes across a spin-Peierls transition should be
expected at the boundary of the Brillouin zone. Optical techniques can only
probe the phonon branch at the $\Gamma$ point ($q=0$). Evidence for a
phonon softening at finite wave vector can therefore only be obtained by
neutron scattering. Nonetheless, one can hope to gain interesting insights
on the temperature dependence of the phonon spectrum, if the dispersion and
the mixing of the branches are not too strong, so that the presence of a
soft mode on the boundaries would result in an overall softening of the
branch it belongs to. Thus, the weak softening of the modes at about 275
\wn~below 100 K might be ascribed to a general red-shift of the $B_{2u}$
branch. It is important to remark, that Raman scattering in \cl~shows
phonon modes with very large fluctuations effects and 20 \% softening
\cite{lmio}. The softening observed in Raman scattering, however, is of
different origin, due to the different polarization of the $A_{g}$ (Raman)
and $B_{2u}$ (IR) modes involved (i.e., along $c$ and $b$ axis,
respectively).

As far as the temperature dependence of the scattering rate ($\Gamma_j$) is
concerned (central panel in Figs. \ref{pa} and \ref{pc}),  along the $b$
axis almost all phonons but the one at 303 \wn~narrow with decreasing
temperature. As in \cl, we associate this narrowing of the phonons along
the $b$ axis to the suppression of low frequency spin fluctuations. This
narrowing is detected in  phonons peaked at lower frequencies, suggesting
that  the spin-fluctuation develops in \bro~at lower energies. On the
contrary, the modes at 417 and 481 \wn~along the $a$ axis show a broadening
with decreasing temperature, while in \cl~the corresponding modes narrow at
low temperature. The remaining modes along $a$ axis do not change the width
with temperature (Fig. \ref{pa}).

The mode strength  $\omega_{pj}$ for the great majority of phonons
decreases at low temperature for both polarizations.  Only  the strength of
the phonons at 481 \wn~for the $a$ axis and at 303 \wn~and 196 \wn~(though
moderately) for the $b$ axis increases. The spectral weight ($SW$), defined
by
\begin{equation}
\label{sw}
SW=\frac{\sum_{j}  w_{pj}^2}{8}=\int\sigma_1(\omega) d\omega,
\end{equation}
displays an overall depletion with decreasing temperature at frequencies below the strong modes at 417 and
275 \wn~for the $a$ and $b$ axis, respectively. Like in \cl, the suppressed
spectral weight is redistributed to higher frequencies.
\begin{figure} [!h]
\begin{center}
    \resizebox*{9.0 cm}{!}{\includegraphics{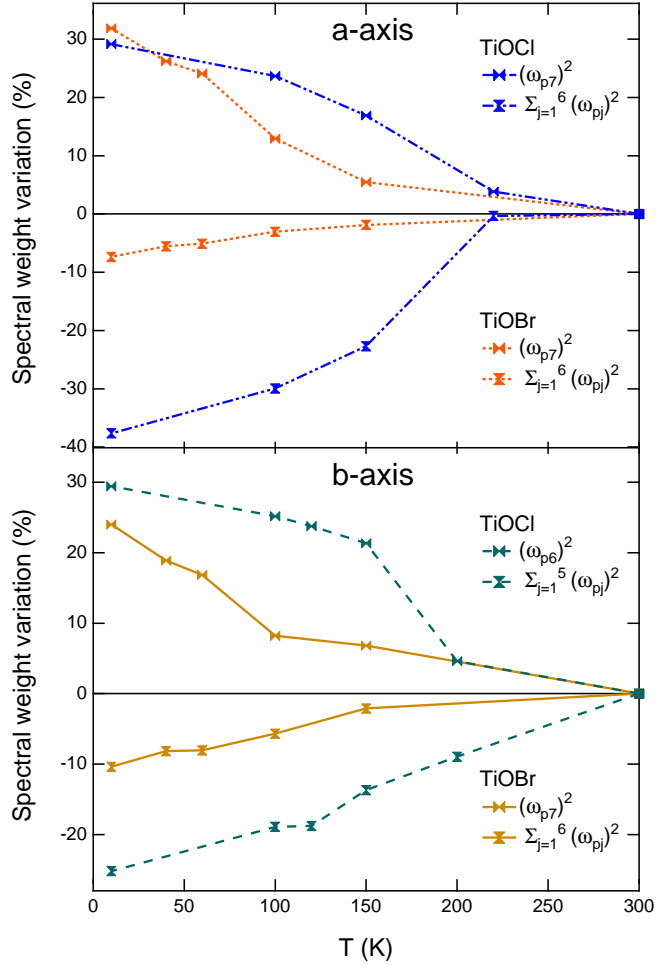}}
\caption{(Color online) Temperature dependence of the spectral weight variation for both polarizations and both samples. The figure highlights the redistribution of spectral weight between low (decreasing $SW$ with decreasing temperature)  and high (increasing $SW$ with decreasing temperature) frequency (see text).}
\label{WP_Var}
\end{center}
\end{figure} 
This is shown in Fig. \ref{WP_Var}, which visualizes the temperature dependence of the spectral
weight redistribution. The decreasing $SW$ is obtained by summing the
squared oscillator strength of each mode at energies smaller than 417 or
275 \wn~ for the $a$  and $b$ axis, respectively. The increasing $SW$ is
encountered in the high frequency tail of the strong IR phonons. An
equivalent analysis (eq. (\ref{sw})) may be performed by integrating
\sig~either from 0 to an appropriate cut-off energy or from such a cut-off
energy up to energies where the \sig~spectra at different temperatures are no more
distinguishable. The cut-off energy can be  chosen in such a way to
differentiate between  energy intervals where a depletion, respectively a
gain in spectral weight has been established (i.e., 417 and 275 \wn~for the $a$ and $b$ axis, respectively). Both analyses show that the
total spectral weight is fully conserved at all temperatures from $\sim1000$ \wn~on.

In \cl, IR optical
and Raman spectroscopy\cite{mio,lmio} displayed a progressive suppression
of $SW$ by decreasing temperature below 200 K. The energy interval, where
the $SW$ depletion occurs in \cl, defines a characteristic energy scale,
attributed to the spin-gap energy $2\Delta_{spin}\approx 430$ K. The
already stressed similarity in the two TiOX compounds leads us to associate
by analogy the depletion of $SW$ in TiOBr, although  less pronounced than
in \cl, to the opening of the spin-gap. The softer onset of the spectral weight depletion (compared to \cl) might indicate the reduced role of lattice dynamics in driving the magnetic  phase transition at T$_{C1}$. The variation of $SW$ in \bro~happens at temperatures (100-150 K)
extending well  above T$_{C1}$ and  T$_{C2}$, pointing out again the
importance of fluctuation effects. Comparing the temperature evolution of
$SW$ in \bro~with that of \cl, one notes that the redistribution of $SW$ in
\bro~develops at lower temperature than in \cl. This goes hand in hand with
the temperature dependence of $\chi$(T), signaling lower critical
temperature for the spin-gap phase.

\section{Conclusion}
We have provided a complete analysis of the IR properties of \bro, with
particular attention on the phonon spectrum. We have shown that mass and
oscillator strength renormalization govern the softening of the phonon
modes of \bro~with respect to those of  \cl.  The temperature dependence of
all relevant parameters determining the IR absorption spectrum develops
over a broad temperature interval extending far above T$_{C1}$. As for \cl,
there is an extended fluctuation regime in \bro~too. The
suppression of spectral weight at low frequency by decreasing temperature hints furthermore to the opening
of a spin-gap.  Moreover, the temperature dependence of the  $SW$ redistribution
mimics the behavior of the spin susceptibility for both compounds, implying a spin-gap phase at lower temperatures in \bro~than \cl. In the TiOX systems the transition at T$_{C1}$ is not a conventional spin-Peierls one. As pointed out for \cl,  the large phonon anomalies \cite{lmio}, the large g-factor line-width\cite{Kataev}, as well as the presence of an intermediate phase between T$_{C2}$ and T$_{C1}$ (Ref. \onlinecite{imai})  hint to competing lattice, spin  and orbital degrees of freedom as the driving force for the  transition.
\begin{acknowledgments}
The authors wish to thank J. M\"uller for technical help, and B. Schlein and A.
Perucchi for fruitful discussions. This work
has been supported by the Swiss National Foundation for the
Scientific Research, INTAS 01-278, DFG SPP1073 and by the MRSEC Program of
the National Science Foundation under award number DMR 02-13282.\end{acknowledgments}

\end{document}